\documentstyle[11pt,newpasp,twoside]{article}
\def\edcomment#1{\iffalse\marginpar{\raggedright\sl#1\/}\else\relax\fi}
\reversemarginpar

\begin{document}
\title{Summary talk: How serious are the problems faced by CDM: cusps,
thin disks, and halo substructure} 
\author{Joel R. Primack}
\affil{Physics Department, University of California, Santa Cruz, CA
95060 USA}

\begin{abstract}
This is a summary of the discussion in Panel I of IAU Symposium 220 on
Dark Matter in Galaxies.  The panel topics and panellists were as
follows -- Cusps Theory: Julio Navarro; Cusps Observations: Erwin de
Blok, Rob Swaters; Substructure Observations: Mario Mateo;
Substructure Theory: James Taylor, Shude Mao.  In addition to the
talks by the panellists and the discussion among them, there was a
good deal of discussion with the audience.  Despite the differences of
opinion expressed I think a consensus emerged, and I try to summarize
it here.
\end{abstract}
    
\section{Introduction}

When the first high-resolution simulations of cold dark matter halos
became available (e.g. Dubinski \& Carlberg 1991), they had a central
density profile approximately $\rho(r) \propto r^{-1}$, which has come
to be known as the central ``cusp.''  It was soon pointed out by
Flores \& Primack (1994) and by Moore (1994) that this central
behavior was inconsistent with the HI observations of dwarf galaxies
that were then becoming available, which suggested that the central
density is roughly constant.  Flores \& Primack (1994) pointed out
that the first cluster lensing observations (Tyson et al. 1990) also
appeared to be inconsistent with a $r^{-1}$ central cusp.  Many
additional rotation curves of low surface brightness (LSB) galaxies
were measured, and they also were claimed to imply that the central
density of these galaxies is rather flat.  It was subsequently pointed
out that the HI observations of galaxies were affected by finite
resolution (``beam smearing''), and that when this was taken into
account the disagreement with simulations is alleviated (see e.g. van
den Bosch et al. 2000, van den Bosch \& Swaters 2001).  More recently,
high resolution H$\alpha$ and CO rotation curves have been obtained
for a number of dwarf and LSB galaxies (e.g. de Blok et al. 2001,
McGaugh et al. 2001, Swaters et al. 2003, Bolatto et al. 2002, and the
papers presented at this Symposium), and there is hope that this will
clarify the situation.

Meanwhile, theorists have done simulations with increasing resolution.
On the basis of simulations with tens of thousands of particles per
dark matter halo, Navarro, Frenk, \& White (1996, 1997, NFW) showed
that halos from galaxy to cluster scales have density profiles that
are described fairly well by the fitting function
$$ \rho_{NFW}(r) \equiv \rho_s (r/r_s)^{-1} (1 + r/r_s)^{-2} $$
and that the halo concentration
$$ c \equiv R_{200}/r_s $$ (where $R_{200}$ is the radius within which
the average halo density is 200 times the critical density) is a
declining function of the mass of the halo.  Subsequently, James
Bullock and Risa Wechsler improved our understanding of halo evolution
in their dissertation research with me, which included analyzing
thousands of dark matter halos in a high-resolution dissipationless
cosmological simulation by Anatoly Klypin and Andrey Kravtsov.
Bullock et al. (2001), defining the (virial) concentration
$$ c_{vir} \equiv R_{vir}/r_s $$ 
(where $R_{vir}$ is the virial radius, see Bullock et al. 2001 and
Bryan \& Norman 1998), found that at fixed halo mass $c_{vir}$ varies
with redshift $z$ as $(1+z)^{-1}$, and developed an approximate
mathematical model that explained the dependence on mass and redshift.
(An alternative model was proposed in Eke et al. 2001, but it appears
to be inconsistent with a recent simulation of small-mass halos by
Colin et al. 2003.)  Wechsler et al. (2002) determined many halo
structural merger trees, and showed that the central scale radius
$r_s$ is typically set during the early phase of a halo's evolution
when its mass is growing rapidly, while $c_{vir}$ subsequently grows
with $R_{vir}$ during the later slow mass accretion phase (see also
Zhao et al. 2003).  Higher resolution simulations with roughly a
million particles per halo gave central density profiles $\rho(r)
\propto r^{\alpha}$ with $\alpha$ as steep as -1.5 (Moore et
al. 1999a, Ghigna et al. 2000, Klypin et al. 2001) although more
recent very high resolution simulations are finding less steep central
profiles with $\alpha \approx -1$ or shallower (Power et al. 2002,
Stoehr et al. 2003, Navarro these proceedings).

\section{Cusps: Observations}

The following remarks attempt to summarize the panel discussion.

\begin{itemize}
\item With the increasing availability of high resolution H$\alpha$
and CO observations, beam smearing is no longer a serious problem.

\item Observational resolution of nearby dwarfs and LSBs has exceeded
the theoretical resolution of simulations.

\item Different observers often agree on $V(r)$ and on the index
$\alpha$ of the same galaxy (where $\rho(r) \propto r^\alpha$).

\item Disagreement remains (Swaters et al. 2003, de Blok et al. 2003,
and these proceedings) on the importance of systematic effects such as
slit width, position, and alignment (all of which bias the slope of
the inner rotation curve to appear shallower than it really is).

\item Two-dimensional data is becoming available at galaxy centers,
including data with exquisite resolution (e.g. Bolatto, these
proceedings).

\item Measured inner slopes $\alpha$ mostly lie in the range 0 to -1.

\item A slope $\alpha = -1.5$ is clearly inconsistent with the observations.

\item The NFW value $\alpha = -1$ is consistent with some galaxies,
possibly consistent with others, and inconsistent with a significant
number.
\end{itemize}

\section{Cusps: Theory}

\begin{itemize}
\item Recent high resolution simulations (see especially Navarro,
these proceedings) have inner slopes inconsistent with $\alpha = -1.5$
(as found by Moore et al. 1999a, Ghigna et al. 2000) and possibly even
with the NFW value $\alpha = -1$.

\item The inner slopes are not converging to any $\alpha$, but appear
to grow shallower at smaller radii (Navarro, these proceedings;
Navarro et al. 2003).  However, the simulations still do not have
sufficient resolution to see whether $\alpha$ actually becomes
shallower than -1 at small radii.

\item Observing test particle motion in a triaxial halo from
$\Lambda$CDM simulations results in a range of 2D velocity profiles
similar to observed ones (Navarro, these proceedings).

\item The mean density $\Delta_{V/2}$ inside the radius $r_{V/2}$
(where the rotation velocity reaches half the maximum observed value)
appears to be somewhat smaller than the $\Lambda$CDM prediction with
$\sigma_8=0.9$, but more consistent with $\Lambda$CDM with $\sigma_8=0.7$
(Alam, Bullock, Weinberg 2002, Zentner \& Bullock 2002) using the
Bullock et al. (2001) model.\footnote{Note however that $\Lambda$CDM with
$\sigma_8=0.9$ is consistent with the early ionization indicated by
WMAP polarization (e.g. Ciardi, Ferrara, \& White 2003), but a tilted
model with $\sigma_8=0.7$ has too little early star formation to
produce the observed optical depth unless there is some sort of exotic
ionization (Somerville, Bullock, \& Livio 2003).}

\item The scatter in the concentrations or $\Delta_{V/2}$ values
indicated by the observations of LSB and dwarf galaxies appears to be
greater than that of the simulations.  (See e.g. Fig. 11 of
Hayashi et. al. 2003.)
\end{itemize}

Although it was not discussed during this panel, I want to mention
some new work on the central cusp problem for clusters.  Sand, Treu,
\& Ellis (2002) measured the density profile in the center of cluster
MS2137-23 with gravitational lensing and velocity dispersion, removed
the stellar contribution with a reasonable M/L, and found
$\rho_{DM}(r) \propto r^\alpha$ with $\alpha \approx -0.35$, in
apparent contradiction to the expected NFW $\alpha=-1$ for CDM.
(See also Treu, these proceedings, and Sand et al. 2003.)
Similar results were found for Abell 2199 by Kelson et al. (2002).
The apparent disagreement with CDM worsens if adiabatic compression of
the dark matter by the infalling baryons is considered (Blumenthal et
al. 1986). However, dynamical friction of the dense galaxies moving in
the smooth background of the cluster dark matter counteracts the
effect of adiabatic compression, and leads to energy transfer from the
galaxies to the dark matter which heats up the central cuspy dark
matter and softens the cusp. N-body simulations (El-Zant et al. 2003)
show that the dark matter distribution can become very shallow with
$\alpha \approx -0.3$ for a cluster like MS2137, in agreement with
observations.

\section{Halo Substructure}

\begin{itemize} 
\item Many fewer satellite galaxies are seen around the Milky Way and
M31 than the number of halos predicted (Kauffmann, White, \&
Guiderdoni 1993; Klypin et al. 1999 $\Lambda$CDM; Moore et al. 1999b
SCDM).
 
\item But for $\Lambda$CDM the discrepancy arises only for satellites
smaller than LMC and SMC, and small satellites are expected to form
stars very inefficiently (Bullock, Kravtsov, \& Weinberg 2000,
Somerville 2002, Benson et al. 2003a).
 
\item Larger ratios of dwarfs/giants are predicted and observed in
cluster cores, where a higher fraction of dwarf halos should have
collapsed before the redshift of reionization, than in lower density
regions like the local group (Tully et al. 2002, Benson et al. 2003a).

\item It is important to check that such dwarf galaxies have expected
properties.  For example, at least some of their stars should be old.

\item Concern: do the predicted radial distributions of halo
substructures agree with observations?  Observed satellites may be
located at smaller radii than the subhalos found in simulations, but
the observational and simulation data sets are still small.

\item ``Milli-lensing'' by DM halo substructure appears to be required
to account for anomalous flux ratios in radio lensing (see
e.g. contributions by Mao, Schechter, and Schneider in these
proceedings).\footnote{Mao \& Schneider 1988 suggested that something
was wrong with the magnification ratios of Q1422.  Metcalf \& Madau
2001 and Chiba 2002 pointed out that dark matter substructure should
be detectible through magnification ratios of strong lenses, 
Metcalf \& Zhao 2002 showed that the measured ratios do not agree with
simple models, and Dalal \& Kochanek 2002 attempted to use this to
measure the density of substructure.  Recent relevant papers include
Schechter \& Wambsganss 2002, Kochanek \& Dalal 2003, Keeton et
al. 2003, and Metcalf et al. 2003.}

\item Concern: is there enough halo substructure in the inner $\sim10$ kpc,
where it appears to be needed to account for such lensing?
\end{itemize}

\section{Disk Thickening}

Another concern is whether halo substructure will thicken disks more
than observed?  Although this was not discussed in the panel, a recent
paper by Benson et al. (2003b) calibrates an analytic model against
N-body simulations to calculate the heating of galactic disks by
infalling satellites.  It concludes that this is a small effect, with
most of the disk thickening due to the gravitational scattering of
stars by molecular clouds.  The observed thicknesses of the disks of
spiral galaxies would then be consistent with the substructure in
dark matter halos predicted by the standard $\Lambda$CDM model.
It would be desirable to check these results with higher-resolution
simulations.

\section{Conclusions}

\begin{itemize}
\item {\bf Cusps}: There has been tremendous progress on observed
velocity fields, and also real progress in improving simulations. {\it
Observed} simulations may agree with observed velocities in galaxy
centers better than seemed likely a few years ago.  But it is
something of a scandal that there is still so little theoretical
understanding of dark matter halo central behavior, although people
are making progress on this problem (e.g. Dekel et al. 2003ab).  It is
likely that poorly understood gastrophysics will turn out to be
relevant (e.g. Combes, these proceedings; Binney, these proceedings).

\item {\bf Substructure}: A challenge appears to be turning into a
triumph for CDM, since it appears that roughly the amount of
substructure predicted by $\Lambda$CDM is required to account for the
number of satellites seen and for the flux anomalies observed in radio
lensing.  The main remaining concern is whether the predicted radial
distribution of substructure agrees with lensing and observed
satellites.  Much work remains to be done to test the theory
quantitatively.

\item {\bf Disk thickening}: Not obviously a problem.
\end{itemize}

There are no better alternatives to $\Lambda$CDM that have yet been
invented.  Those that have been investigated in detail, such as warm
dark matter (WDM) and self-interacting dark matter (SIDM), fare much
worse despite having additional parameters (see e.g. Primack 2003).
Invent new ones, if you can!

\

\noindent {\it Acknowledegments:} I must thank Ken Freeman and Mark
Walker for organizing such an interesting symposium and inviting me to
chair this discussion, Martin Weinberg and many partcipants at the
symposium for helpful discussions, Ben Metcalf for discussions of
gravitational lensing and dark matter substructure, and NSF and NASA
for support at UCSC.

\end{document}